\documentclass[aps,prl,preprint,showpacs,preprintnumbers,amsmath,amssymb]{revtex4}
\usepackage{graphicx,epstopdf,subfigure,natbib}

\begin{document}

\title{Anharmonicity of BaTiO$_3$ single crystals}

\author{Y. L. Wang}
\email{yongli.wang@epfl.ch}
\author{A. K. Tagantsev}
\author{D. Damjanovic}
\author{N. Setter}
\affiliation{Ceramics Laboratory, EPFL-Swiss Federal Institute of Technology,
Lausanne 1015, Switzerland}
\author{V. K. Yarmarkin}
\affiliation{A.F. Ioffe Physical-Technical Institute, St.Petersburg,194021 Russia}
\author{A. I. Sokolov}
\affiliation{Department of Quantum Electronics,
Saint Petersburg Electrotechnical University,
Professor Popov Str. 5,
Saint Petersburg, 197376,
Russia
}
\date{\today}

\begin{abstract}

By analyzing the dielectric non-linearity with the Landau thermodynamic expansion, we find a simple and direct way to assess the importance of the eighth order term. Following this approach, it is demonstrated that the eighth order term is essential for the adequate description of the para/ferroelectric phase transition of BaTiO$_3$. The temperature dependence of the quartic coefficient $\beta$ is accordingly reconsidered and is strongly evidenced by the change of its sign above 165$^oC$. All these findings attest to the anomalously strong polarization anharmonicity of this material, which is unexpected for classical displacive ferroelectrics.

\end{abstract}

\maketitle

Phase transformation phenomena are an important issue in solid state physics.
In understanding these phenomena, both phenomenological and microscopical approaches have been playing comparable roles as one readily finds from the developments of the magnetism, superconductivity, and ferroelectricity.
In particular, considerable attention is paid to the phenomenological Landau-Devonshire theory of ferroelectrics in the modern epoch where more and more advanced techniques of first-principle calculations of ferroelectrics become available.
In the phenomenology of ferroelectrics, the issue recently attracting attention was the role of high-order anharmonic polarization terms in the Landau free-energy expansion \cite{Vanderbilt2001,Gufan2002}. 
In these works, it was demonstrated that the terms containing eighth (or higher) power of polarization, neglected in the traditional phenomenological framework, may be vitally important for the description the phase diagrams of perovskite ferroelectrics.
The physics behind this effect is the need of high-order terms for adequate description of the symmetry of the problem.
Another  demarche in the field has been recently undertaken by Li \textit{et al.} , who have revisited the phenomenological theory  of classical ferroelectric BaTiO$_3$ \cite{Li2005}.
These authors suggested incorporation of the eighth power terms in the Landau expansion of BaTiO$_3$.
In contrast to Refs. \cite{Vanderbilt2001, Gufan2002}, here no symmetry arguments have been involved.
The message by Li \textit{et al.}  was that the Landau expansion  containing eighth power terms (with temperature independent anharmonic coefficients) is as efficient in the description of many properties of BaTiO$_3$  as the traditional sixth order expansion with the temperature dependent anharmonic terms \cite{Bell1984, Bell2001}. 
This conjecture is of the interest since the strong temperature dependence of the anharmonic coefficients in the BaTiO$_3$ Landau expansion is in conflict with the displacive nature of ferroelectricity in the material, at least under the common assumption that the critical fluctuation are weak in this system \cite{Vaks1970, Sokolov2002}.
Interestingly, in their analysis of the Landau potential using first principle calculations, Iniguez \textit{et al.} obtained that the sixth order expansion accounts for the main features of BaTiO$_3$ phase diagram, but that all coefficients in the expansion have nontrivial temperature dependence \cite{Iniguez2001}.

Together, these results pose two questions. First, are higher than sixth power terms in the BaTiO$_3$ Landau expansion needed for the correct physical description and are not just a matter of convenience? Second, does the incorporation of such terms enable elimination of the strong temperature dependence of the anharmonic coefficients in the BaTiO$_3$ Landau expansion? The present paper addresses these questions using the authors' experimental data and those available in the literature to unambiguously obtain a positive answer to the first question and a negative to the second.
   
\begin{figure}
\begin{center}
\includegraphics[width=6cm]{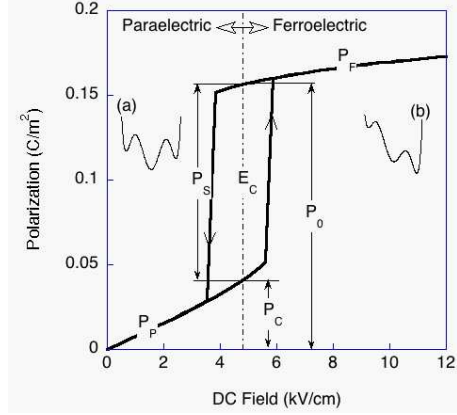}
\caption{(Color online) Schematic of the phase transition induced by an electrical field applied along [001] axis above \textit{T$_C$} for BaTiO$_3$ single crystal. In paraelectric phase the polarization varies with field below $P_C$, while in ferroelectric phase the polarization is dominated by $P_0=P_C+P_S$. Curves (a) and (b) schematize the typical profiles of the Gibbs potential in paraelectric and ferroelectric phase, respectively.}
\label{double hysteresis loop}
\end{center}
\end{figure}

To demonstrate the crucial role of the eighth-power term in the Landau expansion we will address the dielectric non-linearity of BaTiO$_3$ with respect to the electric field $E$ applied along the $[001]$ axis, in the vicinity of the cubic/tetragonal phase transition temperature.
In this case only $[001]$ component $P$ of the polarization is involved, so that the Landau expansion of the Gibbs potential with the eighth-power term reads
\begin{equation}
\label{Gibbs8}
\triangle G=    \frac{1}{2} \alpha P^{2}+
            \frac{1}{4} \beta P^{4}+
            \frac{1}{6} \gamma P^{6}+
            \frac{1}{8} \delta P^{8}
\end{equation}
Accordingly we have for the equation of state and $c$-axis permittivity $\epsilon_c$: 
\begin{equation}
\label{E8}
E=           \alpha P+
             \beta P^{3}+
             \gamma P^{5}+
             \delta P^{7}
\end{equation}
\begin{equation}
\label{lambda8}
\epsilon_c^{-1}=
             \alpha +
            3 \beta P^{2}+
            5 \gamma P^{4}+
            7 \delta P^{6}
\end{equation}
 These equations provide description of the non-linear dielectric response of BaTiO$_3$ in terms of the eighth-power expansion.
 One can compare this description with that in terms of the sixth-power expansion 
\begin{equation}
\label{E6}
E=           \alpha P+
             \beta^{\prime} P^{3}+
             \gamma^{\prime} P^{5}
\end{equation}
\begin{equation}
\label{lambda6}
\epsilon_c^{-1}=
             \alpha +
            3 \beta^{\prime} P^{2}+
            5 \gamma^{\prime} P^{4}
\end{equation} 
which has been traditionally used in the field. 
These descriptions are clearly different. 
Formally, as one can readily check, the situation described by the expansion Eq. \eqref{E8} with the polarization-independent coefficients $\beta, \gamma$, and $\delta$ corresponds to  polarization-dependent coefficients $\beta^{\prime}$ and $\gamma^{\prime}$ in Eq. \eqref{E6}: 
\begin{equation}
\label{beta'}
\beta^{\prime} =\beta-\delta P^{4}, \gamma^{\prime} =\gamma+2\delta P^{2}
\end{equation}
In general, the consideration of the polarization-dependent anharmonic coefficients in the Landau expansion has no physical sense.
However, if a not too wide interval of polarization variation is experimentally addressed with a finite measurement accuracy, coefficients $\beta^{\prime}$ and $\gamma^{\prime}$ defined by Eq. \eqref{beta'} can be considered as polarization-independent when fitting the dielectric non-linearity. In other words, serious problems may arise in situations when a large variation of polarization is involved, e.g. the "jump" of the polarization on crossing the first order phase transition. To verify this point we consider the case just above $T_C$, where the ferroelectric phase can be induced by an electrical field. As is schematized by the $E\geqslant0$ segment of the "double hysteresis loop" in Fig. \ref{double hysteresis loop}, the polarization "jumps" at the critical field $E_C$, whereas below or above this value the polarization varies slowly. 
%
%Figure I -----------------------------------------------------------------------------
%Figure I -----------------------------------------------------------------------------
%
At electric fields corresponding to the paraelectric phase, the contributions of the polarization dependent terms can be estimated from Eq. \eqref{beta'} to be not more than $\delta {P_C}^4$ and 2$\delta {P_C}^2$, where $P_C$ is the maximal values of the polarization in this phase.
Using the $\delta$ value of 2.9$ \times 10^{11}$ $\mathrm{Vm^{13}C^{-7}}$ \cite{Li2005}, and the approximate value of $P_C$ in our measurements ($\sim$0.04 $\mathrm{C/m^{2}}$), the magnitudes of the correcting terms in Eq.\eqref{beta'} are around 8$ \times 10^5$ $\mathrm{Vm^5C^{-3}}$ and 8$ \times 10^8$ $ \mathrm{Vm^9C^{-5}}$ for $\beta$ and $\gamma$, respectively.
In comparison with the typical values of $\beta$ (-6$\sim$9$ \times 10^8$ $\mathrm{Vm^5C^{-3}}$) and $\gamma$ (1$\sim$2$ \times 10^{10}$ $\mathrm{Vm^9C^{-5}}$), these corrections are negligible to within at least of a few percent. 

At electric fields corresponding to the ferroelectric phase, the polarization is much larger than in the paraelectric phase.
Provided that the interval of the polarization variation addressed is much smaller than the minimal polarization in the induced ferroelectric phase, $P_0$, when evaluating $\beta^{\prime}$ and $\gamma^{\prime}$, one can take into account only the contribution of $P_0$ in the correcting terms in Eq. \eqref{beta'}, leading to the polarization independent $\beta^{\prime}$ and $\gamma^{\prime}$
\begin{equation}
\label{betaT}
\beta^{\prime} \approx \beta-\delta {(P_0)}^4, \gamma^{\prime} \approx \gamma+2\delta {(P_0)}^2.
\end{equation}
Using the typical value of $P_0$ ($\sim$0.16 $ \mathrm{C/m^2}$), the second terms in these relations can be estimated as 2$ \times 10^8$ $\mathrm{Vm^5C^{-3}}$ and 1.5$ \times 10^{10}$ $\mathrm{Vm^9C^{-5}}$ for $\beta$ and $\gamma$, respectively.
These corrections are comparable with the typical values of the corresponding coefficients given above and cannot be neglected.

Thus, from the above analysis, we conclude that if the eighth-power term in the Landau expansion exists having a value of the order of that estimated by Li \textit{et al.}  \cite{Li2005}, the anharmonic polarization coefficients $\beta^{\prime}$ and $\gamma^{\prime}$ of BaTiO$_3$ estimated in terms of the sixth-power expansion should exhibit appreciable jumps on crossing the field induced phase transition.
Such "phase sensitivity" of the nonlinear coefficients of the sixth order expansion enables the detection of  the eighth-power term, suggesting a simple and direct way to assess the anharmonicity of BaTiO$_3$.  

Our experiments unambiguously detect this phase sensitivity.
We have performed precise measurements of the dc field dependence of the dielectric permittivity of a BaTiO$_3$ single crystal slightly above $T_C$ along [001] direction.
An undoped BaTiO$_3$ single crystal (commercially available, GB group, Inc.) is cut and polished into a 2$\times$2$\times$0.3 mm$^3$ plate, with the main surface perpendicular to [001].
The two main sides are then entirely coated with platinum electrode to form a capacitor.
A computer controlled setup, consisting of a temperature chamber, a high voltage power supply, and a precision LCR meter, was used to control or measure the temperature, the bias voltage, and dielectric properties (10 kHz), respectively.

%Figure 2 -----------------------------------------------------------------------------
\begin{figure}
\begin{center}
\includegraphics[width=6cm]{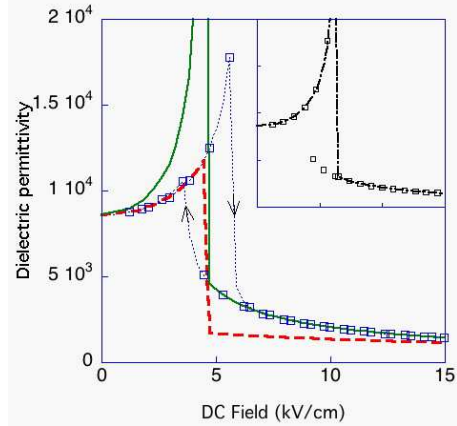}               
\caption{(Color online) Field dependence of the dielectric permittivity for BaTiO$_3$ single crystal at 135$^oC$. The open squares represent the measured values, and the dotted line is to guide the eyes. Remarkably different $\beta$ are obtained for either phase when fitting with Eq. \eqref{E6} and \eqref{lambda6}, as listed in Table \ref{tab1}. The dashed and solid lines are calculated with the coefficients derived from the paraelectric and ferroelectric data, respectively. The "ferroelectric" coefficients have serious problem in describing the dielectric properties of the paraelectric phase, and vice versa. The inset shows much better fit quality of an eighth order expansion for both phases.}
 \label{fig2}
\end{center}
\end{figure}
%Figure 2 -----------------------------------------------------------------------------

Figure \ref{fig2} shows the experimental data measured at 135$\mathrm{^oC}$, which is around 4 $\mathrm{^oC}$ higher than $T_C$. Little hysteresis effect related to the relaxation phenomena is found within either phase, indicating the high insulating quality of the crystal and the space charge free state of the crystal-electrode interface \cite{Triebwasser1960}.
The sixth order expansions, Eq. \eqref{E6} and \eqref{lambda6}, are used to fit the curves with the common value of $\alpha$, and remarkably different $\beta$ are obtained for either phase, as listed in Table \ref{tab1}.
As for $\gamma$, we cannot determine it only from the nonlinear dielectric behavior of the paraelectric phase. The position of $E_C$ is used to obtain an estimate, which remarkably differs from that derived in the ferroelectric phase.    
The ferroelectric-phase coefficients have a serious problem in describing the dielectric properties of the paraelectric phase, and vice versa.
As illustrated in the inset of Fig. \ref{fig2}, the eighth order expansion with a common set of coefficients enables a good fit in both phases at the same time.
As-estimated values of $\beta$, $\gamma$, and $\delta$, agree well with the set suggested by Li \textit{et al.}  \cite{Li2005}.
%Table I -----------------------------------------------------------------------------
\begingroup
\squeezetable
\begin{table}
\caption{Nonlinear coefficients obtained from the field dependence of dielectric permittivity shown in Fig. \ref{fig2}.}\label{tab1}
\begin{tabular}{l l l l l}
\hline
\hline
Order & Phase(s) & $-\beta\mathrm{(Vm^5C^{-3})}$ & $\gamma\mathrm{(Vm^9C^{-5})}$ & $\delta\mathrm{(Vm^{13}C^{-7})}$  \\
\hline
6 & Paraelectric & $8.6\times 10^8$ & $1\times 10^{10}$ \footnotemark & 0 \\
6 & Ferroelectric & $18.7\times 10^8$ & $8.2\times 10^{10}$ & 0 \\
8 & Both & $8.6\times 10^8$ & $0.8\times 10^{10}$ & $3.4\times10^{11}$ \\
\hline
\hline
\end{tabular}
\footnotetext{The value of $\gamma$ cannot be determined only by fitting the paraelectric branch. The present value is estimated by combining the dielectric nonlinearity and the position of the critical field $E_C$.}
\end{table} 
\endgroup
%Table I-----------------------------------------------------------------------------

    Additional evidence for the eighth order term can be obtained using the dielectric permittivity values at $T_C$, where the phase transition is induced by temperature in the absence of external field.
Taking into account that at $T_C$, $E=0$ and $\Delta G=0$, Eqs.\eqref{E8} and \eqref{lambda8} lead to a simple relation for the variation of the dielectric permittivity: $R=\frac{\epsilon_{c, C}}{\epsilon_{c, T}}=4+\frac{\delta {P_S}^6}{\alpha}$, where $\epsilon_{c, C}$ and $\epsilon_{c, T}$ stand for lattice permittivity  along [001] at $T_C$ in the cubic and tetragonal phases, respectively. 
One sees that, if $\delta$ were zero or small enough, the ratio $R$ would be close to 4 \cite{adiabatic}.
In Table \ref{tab2}, our data for $R$ and those available in the literature are summarized.
Apparently most of the values are remarkably larger than 4.
Considering the possible experimental inaccuracy in determining the lattice permittivity in the ferroelectric phase due to not fully eliminated domain contribution, the real value of $R$ is expected to be even larger than the presented values.
The deviation of $R$ from 4 provides additional evidence for the important role of the eighth order term.  
%Table II -----------------------------------------------------------------------------
\begin{table}
\caption{Dielectric permittivity of BaTiO$_3$ single crystal at the phase transition temperature, $T_C$.}\label{tab2}
  \begin{tabular}{l l l l}
\hline
\hline
Authors, Ref. & $\epsilon_{c, C}$ & $\epsilon_{c, T}$ & \textit{R}  \\
\hline
Merz, \cite{Merz1953} & 14,200 & 2,300 & 6.2 \\
Drougard, \cite{Drougard1954} & 16,000 & 2,200 & 7.3 \\
Meyerhofer, \cite{Meyerhofer1958} & 16,500 & 5,000 & 3.3 \\
Johnson, \cite{Johnson1965} & 9,130 & 1,500 & 6.1 \\
This work & 11,400 & 2,200 & 5.2 \\
\hline
\hline
\end{tabular}
\end{table} 
%Table II-----------------------------------------------------------------------------

\begin{figure}
\begin{center}
\includegraphics[width=6cm]{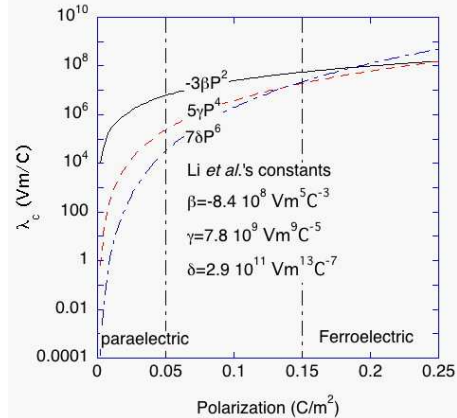}    
\caption{(Color online) Contributions of the various terms to the inverse dielectric permittivity, following Li \textit{et al.} 's constants. In the low polarization regime, typically in paraelectric phase, the nonlinear contribution is dominated by the quartic term. In the high polarization region, typically in ferroelectric phase, the contribution of the eighth order term is comparable to or even larger than that of the sixth order term.}
  \label{fig3}
\end{center}
\end{figure}

To elucidate more clearly the contributions of the various terms to the inverse dielectric permittivity, these are plotted as functions of polarization in Fig. \ref{fig3}, following Li \textit{et al.} 's constants.
It is seen that, in the low polarization regime, typically in the paraelectric phase, the quartic term dominates the whole nonlinear contribution and the dielectric nonlinearity can therefore be well fitted even by fourth order expansion. In the high polarization regime, typically in the ferroelectric phase, the contribution of the eighth order term becomes comparable to or even larger than that of the sixth order term. 
%
%Figure 3 -----------------------------------------------------------------------------
%Figure 3 -----------------------------------------------------------------------------
%

\begin{figure}
\begin{center}
\includegraphics[width=6cm]{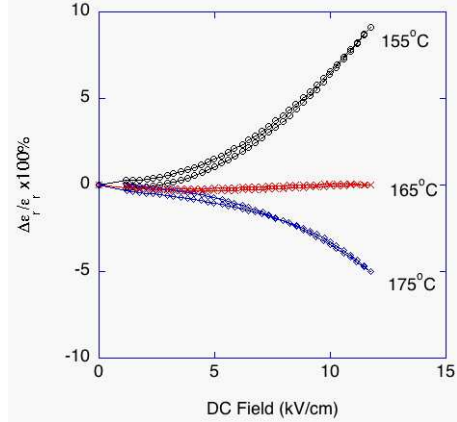}
\caption{(Color online) Field dependence of dielectric permittivity at three representative temperatures, evidencing the reversal of the sign of $\beta$ at around 165$^oC$}.
\label{fig4}
\end{center}
\end{figure}

Thus, we have shown that the eighth-power term is vitally important in the thermodynamics of BaTiO$_3$, supporting the first conjecture by  Li \textit{et al.} .
However, their  second conjecture, that the eighth-power term can be introduced to avoid the use of temperature dependent anharmonic coefficients in the BaTiO$_3$ Landau expansion, remains unchecked.
The direct way to do that is to analyze the dielectric non-linearity in the paraelectric phase in terms of the full expansion.
As mentioned above, in the paraelectric phase, the dielectric nonlinearity is predominately controlled by the quartic coefficient $\beta$.
For this reason, the experimental results on a substantial temperature dependence of $\beta^{\prime}$ of the sixth-power expansion in the paraelectric phase \cite{Drougard1955, Meyerhofer1958, Kaczmarek1965, Gonzalo1971} strongly suggests that $\beta$ coefficient from the full expansion is also temperature dependent.
However the aforementioned experimental data are rather scattered.
To clarify this point, we measured the dielectric permittivity as a function of the field at a series of temperature points to find a reversal of the sign of the derivative at some 35 K above $T_C$ (Fig. \ref{fig4}), implying a change of sign of $\beta$ \cite{positive beta}. 
Such behavior  is clearly impossible on the assumption of temperature independent $\beta$, even in terms of the eighth-power expansion.
From this, we can definitely conclude that the thermodynamic framework with a temperature independent $\beta$ coefficient does not provide an adequate thermodynamic description of BaTiO$_3$ even  in terms of the eighth-power Landau expansion.      
Thus, our results do not support the second conjecture by  Li \textit{et al.} . 
%
%Figure 4 -----------------------------------------------------------------------------
%Figure 4 -----------------------------------------------------------------------------

In conclusion, analyzing  the non-linear and linear dielectric data on BaTiO$_3$ single crystal at temperatures close to ferroelectric/paraelectric phase transition we have demonstrated that the eighth-power anharmonic polarization term plays an essential role in the thermodynamics of the material. Our result justifies the corresponding conjecture by Li \textit{et al.} \cite{Li2005}.
At the same time, based on the result on the temperature dependence of the dielectric non-linearity in the paraelectric phase, we have demonstrated that even the eighth-power Landau expansion should contain temperature dependent anharmonic polarization terms in order to provide an adequate thermodynamic description of the material. All these findings attests to anomalously strong polarization anharmonicity of BaTiO$_3$, which is unexpected for classical displacive ferroelectrics.    
%
% Acknowledgments ------------------------------------------------------------
\begin{acknowledgments}
This project was supported in part by the Swiss National Science Foundation. Additional support from MIND-European net-work on piezoelectricity is gratefully acknowledged. A. I. Sokolov acknowledges the support of the Russian Foundation for Basic Research under Grant No. 04-02-16189 and of Russian Ministry of Science and Education within Project RNP.2.1.2.7083. 
\end{acknowledgments}
% Acknowledgments ------------------------------------------------------------

%Reference -----------------------------------------------------------------------

%Reference -----------------------------------------------------------------------


\begin{thebibliography}{999}
\bibitem{positive beta} The decreasing field dependence of dielectric permittivity might be attributed to the effect of the higher order terms in case of a zero $\beta$. We have estimated this effect with the typical value of $\gamma$ and found that this effect is negligibly small at 175$^oC$.
\end{thebibliography}
\end{document}